\documentclass{article}
\usepackage{graphicx}
\newcommand{\Max}{\mbox{Max}}
\begin{document}
\baselineskip 20pt
\vglue 2cm
\centerline{\Large {\bf The B\"acklund transformation equations}}
\centerline{\Large {\bf for the ultradiscrete KP equation}}

\vglue 4cm
\centerline{Nobuhiko SHINZAWA$^{\star}$\footnote{
nshinzawa@mpdt.math.waseda.ac.jp}
 \quad and \quad Ryogo HIROTA$^{\dagger}$\footnote{
roy@spn1.speednet.ne.jp}}
\vglue 1cm
\centerline{{\it $\star$ Department of Mathematical Sciences, School of Science And Engineering, Waseda University}}
\centerline{{\it $\dagger$ Profesor emeritus, Waseda University}}
\centerline{{\it Ohkubo 1-1, Shinjuku, Tokyo 192-03, Japan}}
\vglue 6cm

\begin{abstract}
The B\"acklund transformation for the ultradiscrete KP equation 
is proposed. 
An algorithm to eliminate variables from the ultradiscrete linear equations is
proposed.
The consistency condition for the B\"acklund transformation equations 
is obtained via the algorithm.
\end{abstract}

\clearpage


\section{Introduction}
The B\"acklund transformation equations play an important role 
in the study of integrable equations \cite{Rogers}. 
They generate the B\"acklund transformation which allows one to obtain the
exact N soliton solution of the nonlinear equation,
and also provide the Lax pair.
The integrable equation itself is given as a consistency condition
of the B\"acklund transformation equations in these schemes.

On the one hand, the cellular automaton models which have 
exact N soliton solution were found in the last decade \cite{Takahashi}.
In many cases, these soliton cellular automaton models are
obtained as a special limit, known as the ultradiscrete limit,
of the discrete integrable equations.
Typically, the limiting procedure simply replace the summation
by the Max operator and the product by the summation.
Despite of this simple correspondence, 
this procedure causes some ambiguities concerning to the Max operator.
For example, we can remove the variable $x$ from the equation $a+x = b+x$,
however, we can not remove the variable $x$ from
\begin{eqnarray*}
\mbox{Max}(a,x) = \mbox{Max}(b,x) .
\end{eqnarray*}
These type of difficulties prevent us to consider the consistency condition
of the B\"acklund transformation equations 
for the soliton cellular automaton model.

In this paper, we consider the ultradiscrete KP equation and its 
B\"acklund transformation equations, 
since it reduces to many soliton cellular automaton models.
The ultradiscrete KP equation was first proposed in the paper \cite{udkp}, 
as the unified equation of a class of
generalized Box and Ball systems.
It also contains the ultradiscrete Toda equation \cite{Takahashi}, etc.

We show the consistency condition of the B\"acklund transformation equations
of the ultradiscrete KP equation as follows.
First, we propose an extension of the B\"acklund transformation equations
of the discrete KP equation \cite{Shinzawa}.
An important aspect of the extended B\"acklund transformation equations 
is that, it can be expressed as the vanishing of the product of a matrix
and a vector.
The consistency condition becomes the vanishing of the determinant of
the matrix.
This will be shown in the next section.
Second, in section 3, we show the algorithm to eliminate N variables from 
N ultradiscrete linear equations, before considering the
ultradiscretization of the discrete KP equation.
Finally, in section 4, we derive the  B\"acklund transformation equations of
the ultradiscrete KP equation,
starting from the extended B\"acklund transformation equations of section 2.
By applying the algorithm of section 3,
we derive the consistency condition of the B\"acklund transformation equations.
Some examples of soliton solutions are also shown.


\section{B\"acklund transformation equation for the discrete KP equation}

Let us begin with writing down the discrete KP equation for a function $f$
of three discrete variables $p,q,r$ \cite{Hirota} \cite{KP}, 
\begin{eqnarray}
z_{p}z_{qr}f_{p}f_{qr}+z_{q}z_{rp}f_{q}f_{pr}+z_{r}z_{pq}f_{r}f_{pq}=0 .
\label{discreteKP}
\end{eqnarray}
Here, $z_{i}$'s are arbitrary complex constants and $z_{ij}=z_{i}-z_{j}$.
The lower subscripts of $f$ indicate to increase the corresponding
variables by one,
\begin{eqnarray}
f_{p}=f(p+1,q,r) , \nonumber \\
f_{pq}=f(p+1,q+1,r) .
\end{eqnarray}
Despite of the simplicity, this equation plays an important role 
in the study of the integrable equations.
It reduces to many integrable equations known in the physics and mathematics
 \cite{Hirota}, 
and is satisfied by the transfer matrices of a class of solvable 
lattice models \cite{McCoy},\cite{Kuniba},\cite{Krichever}
, and is also satisfied by the amplitude of a string model \cite{Saito2}. 

The B\"acklund transformation equations for the discrete KP equation
are a system of linear equations
\begin{eqnarray}
z_{r}f_{r}g_{q}-z_{q}f_{q}g_{r}+z_{qr}f_{qr}g=0 , \nonumber\\
z_{r}f_{r}g_{p}-z_{p}f_{p}g_{r}+z_{pr}f_{pr}g=0 .
\label{backlund}
\end{eqnarray}
If we substitute a solution of the discrete KP equation
into the function $f$ in this equation, then we can show that 
there exist nontrivial solutions $g$, which satisfy the discrete KP equation.
The compatibility condition of these two equations is in the central
of this scheme.
It guarantees the existence of the solution for these equations,
and it also guarantees that the solution, $g$ also satisfies the discrete KP
equation.
See \cite{Hirota} and \cite{Saito} for example.
However, the consistency condition does not work 
for the ultradiscretized system,
because of the ambiguity of the Max operator stated above.

In this paper, we extend 
the B\"acklund transformation equations (\ref{backlund}), 
by considering the additional bilinear equations 
deduced from the discrete KP equation and 
its B\"acklund transformation equations (\ref{backlund}).
The additional bilinear equations provide another expression of
the B\"acklund transformation equations.

Let us rewrite the B\"acklund transformation equations,
by dividing them by $g$ and move the terms other than $f_{qr}$ and $f_{pr}$
to the right hand side
\begin{eqnarray}
f_{qr}=\frac{-z_{r}f_{r}g_{q}+z_{q}f_{q}g_{r}}{z_{qr}g}, \nonumber\\
f_{pr}=\frac{-z_{r}f_{r}g_{p}+z_{p}f_{p}g_{r}}{z_{pr}g} .
\end{eqnarray}
Substituting these expressions of $f_{qr}$ and $f_{pr}$ into the
discrete KP equation, we can obtain an additional bilinear equation
\begin{eqnarray}
z_{p}f_{p}g_{q}+z_{q}f_{q}g_{p}-z_{pq}f_{pq}g=0 .
\label{additional1}
\end{eqnarray}
Similarly, substituting $f_{p}$ and $f_{q}$ 
appearing in the equations (\ref{backlund}) into the discrete KP equation,
we obtain another bilinear equation
\begin{eqnarray}
z_{pq}f_{pq}g_{r}+z_{qr}f_{qr}g_{p}-z_{pr}f_{pr}g_{q}=0 .
\label{additional2}
\end{eqnarray}
Notice that the equations (\ref{backlund}), (\ref{additional1}) 
and (\ref{additional2}) are linearly independent,
unless the function $f$ satisfies the discrete KP equation.
These four equations are the extension of 
the B\"acklund transformation equations, which we employ in this paper.

An important aspect of the extended B\"acklund transformation equations
is that they can be summarized into the product of a matrix and a vector.
Indeed, the extended B\"acklund transformation equations consist of
four equations and the number of $g$ appearing in the same equations 
(i.e $g,g_{p},g_{q},g_{r}$) is four, thus, we can write down the equations
by using a 4 $\times$ 4 matrix and a 4 component vector
\begin{eqnarray}
\left[\begin{array}{cccc}
0&-z_{r}f_{r}&z_{q}f_{q}&z_{rq}f_{rq}\\
z_{r}f_{r}&0&-z_{p}f_{p}&z_{pr}f_{pr}\\
-z_{q}f_{q}&z_{p}f_{p}&0&z_{qp}f_{pq}\\
z_{qr}f_{qr}&z_{rp}f_{pr}&z_{pq}f_{pq}&0\\
\end{array}
\right]
\left[\begin{array}{cccc}
g_{p}\\
g_{q}\\
g_{r}\\
g
\end{array}\right]=0 .
\end{eqnarray}
The consistency condition for these equations is the
 existence of the nontrivial solution, $g$
and it is nothing but the vanishing of the determinant of this matrix.
In this case, the matrix is antisymmetric matrix, thus, 
we can use the fact that the determinant of the antisymmetric matrix becomes
the square of a pfaffian.
The vanishing of the pfaffian becomes just the discrete KP equation.

Since the consistency condition is very different 
from the consistency condition for the original B\"acklund transformation
equations,
there is a chance to avoid the difficulty of the Max operator.


\section{Ultradiscrete linear equations}
Of course, we can remove the Max operator from the ultradiscrete B\"acklund
transformation equations, by dividing it into cases.
However, it is complicated and requires the individual consideration.
Instead to do so, we study the generic consistency condition
for the ultradiscrete linear equations.
We show an algorithm to eliminate variables, starting from the 
consideration of two linear equations.

Before proceeding to the generic cases,
let us consider the following two examples of two ultradiscrete 
linear equations.
Our first example is the following couple of equations
\begin{eqnarray}
\tilde{a}+x=\Max(a+x,b+y) , \nonumber\\
\tilde{c}+x=\Max(c+x,d+y) .
\label{example1}
\end{eqnarray}
Here, $a,b,c,d,\tilde{a},\tilde{c}$ are the constant coefficients, 
and $x,y$ are the variables of these two equations.
We can eliminate these variables, by considering the
 trivial identity
\begin{eqnarray}
\Max(\Max(a+x,b+y)+d,c+b+x)=\Max(a+d+x,\Max(c+x,d+y)+b) .
\end{eqnarray}
Substituting the first and second equations of (\ref{example1}) into
the left hand side and the right hand side of this equation respectively,
we obtain the relation
\begin{eqnarray}
\Max(\tilde{a}+d,c+b) = \Max(a+d,\tilde{c}+b) .
\end{eqnarray}
This relation should be satisfied, 
if there are finite solutions, $x$ and $y$.
Thus, we can eliminate all variables, 
without dividing Max operators into the cases.

Our second example is the following couple of linear equations
\begin{eqnarray}
\tilde{a}+x = \Max(a+x,b+y) , \nonumber\\
\tilde{d}+y = \Max(c+x,d+y).
\end{eqnarray}
We can also eliminate all variables from these equations without dividing them
into cases.
Indeed, adding these two equations and applying these two equations
to the left hand side once again, we obtain the relations
\begin{eqnarray}
&&\tilde{a}+\tilde{d}+x+y \nonumber\\
&&=\Max(a+c+2x,a+d+x+y,b+c+x+y,b+d+2y) \nonumber\\
&&=\Max(a+\tilde{d}+x+y,\tilde{a}+d+x+y,b+c+x+y) .
\end{eqnarray}
We can eliminate all variables, by subtracting the $x+y$ from the first part
and the third part of these equations,
\begin{eqnarray}
\Max(\tilde{a}+d,a+\tilde{d},b+c) = \tilde{a}+\tilde{d} .
\end{eqnarray}

In the general two linear equations cases, 
we can not eliminate all variables without dividing them into cases.
However, previous two examples make the consideration simple.
The most general ultradiscrete two linear equations 
for two variables $x$ and $y$ are written as 
\begin{eqnarray}
\Max(\tilde{a}+x,\tilde{b}+y)=\Max(a+x,b+y), \nonumber\\
\Max(\tilde{c}+x,\tilde{d}+y)=\Max(c+x,d+y) .
\label{generictwo}
\end{eqnarray}
For example, if $\tilde{a}+x$ in the left hand side of the first equation
is larger than $\tilde{b}+y$ and $\tilde{c}+x$ in second equation is
larger than $\tilde{d}+y$, these equations become as
\begin{eqnarray}
\tilde{a}+x = \Max(a+x,b+y), \nonumber\\
\tilde{c}+x = \Max(c+x,d+y) .
\end{eqnarray}
This is the case of our first example.
By applying the result of the first example, we can eliminate all variables
from these equations.
\begin{eqnarray}
\Max(\tilde{a}+d,c+b)=\Max(a+d,\tilde{c}+b)
\end{eqnarray}
Applying the similar argument to the rest of the three cases
\begin{eqnarray}
&&\tilde{a}+x \ge \tilde{b}+y \mbox{ and } \tilde{c}+x \ge \tilde{d}+y 
\quad \Rightarrow \quad \Max(\tilde{a}+d,c+b)=\Max(a+d,\tilde{c}+b) ,\nonumber\\
&& \tilde{a}+x \le \tilde{b}+y \mbox{ and } \tilde{c}+x \le \tilde{d}+y  
\quad \Rightarrow \quad
\Max(\tilde{b}+c,d+a)=\Max(b+c,\tilde{d}+a) ,\nonumber\\
&& \tilde{a}+x \ge \tilde{b}+y \mbox{ and } \tilde{c}+x \le \tilde{d}+y
\quad \Rightarrow \quad
\tilde{a}+\tilde{d}=\Max(\tilde{a}+d,a+\tilde{d},b+c) ,\nonumber\\
&&\tilde{a}+x \le \tilde{b}+y \mbox{ and } \tilde{c}+x \ge \tilde{d}+y 
\quad \Rightarrow \quad
\tilde{b}+\tilde{c}=\Max(\tilde{b}+c,b+\tilde{c},a+d) \quad , 
\end{eqnarray}
and summarizing these results in one equation, we can obtain the relation
\begin{eqnarray}
\Max(a+d, \tilde{a}+\tilde{d},b+\tilde{c},\tilde{b}+c)=
\Max(\tilde{a}+d,a+\tilde{d},b+c,\tilde{b}+\tilde{c}) .
\label{consistency_two}
\end{eqnarray}
One can easily confirm that this relation is satisfied in all
four cases.
This relation should be satisfied, if there are finite solution, $x$ and 
$y$ for the generic two linear equations (\ref{generictwo}).

On the one hand, we can rewrite the equations (\ref{generictwo}) as
\begin{eqnarray}
\Max(\tilde{a}+z,\tilde{b})=\Max(a+z,b) , \nonumber\\
\Max(\tilde{c}+z,\tilde{d})=\Max(c+z,d) .
\end{eqnarray}
Here, we subtract $y$ from equations (\ref{generictwo}), 
and replace $x-y$ by $z$.
Apparently, if there is finite $z$ which satisfies these equations,
the relation (\ref{consistency_two}) should be satisfied.
Since, $a,b,c,d,\tilde{a},\tilde{b},\tilde{c},\tilde{d}$ are 
arbitrary constants, we can eliminate one variable from two linear equations
of any number of variables.
Applying this reduction recursively,
we can reduce N linear equations for N variables to 
 2 linear equations
for 2 variables, and the rest of 2 variables can be eliminated from 
 two linear equations.
So that, we can eliminate N variables from N linear equations.

In the next section, we consider the consistency condition of the B\"acklund
transformation equations for the ultradiscrete KP equation 
by using this algorithm.


\section{The B\"acklund transformation equations for the ultradiscrete 
KP equation}

Let us write the discrete KP equation, once again.
It was the following equation
\begin{eqnarray}
z_{p}z_{qr}f_{p}f_{qr}+z_{r}z_{pq}f_{r}f_{pq}=z_{q}z_{pr}f_{q}f_{pr} .
\label{discreteKP2}
\end{eqnarray}
Here, $f$ is a complex valued function in the original discrete KP equation,
but we restrict the value to positive real numbers, 
to take the ultradiscrete limit.
We also restrict the values of $z_{i}$'s to real numbers, 
and assume $z_{p}>z_{q}>z_{r}$ since it does not lose the generality.
Notice that in this choice of coefficients, both of the left hand side
and the right hand side of the equation (\ref{discreteKP2}) become positive.

The ultradiscrete limit of the discrete KP equation is taken as follows.
First, we transform the dependent variable $f$ and 
the coefficients $z_{i},z_{ij}$ 
into $F$ and $Z_{i},Z_{ij}$, as 
\begin{eqnarray}
f=\exp(\frac{F}{\epsilon})  \quad , \quad 
z_{i}=\exp(\frac{Z_{i}}{\epsilon}) \quad , \quad
z_{ij}=\exp(\frac{Z_{ij}}{\epsilon})\quad .
\label{ultradiscritization} 
\end{eqnarray}
Here, $\epsilon$ is a positive real parameter.
Second, we apply the log function to both side of 
the equation (\ref{discreteKP2}) after
multiplication of $\epsilon$,
and take the $\epsilon \rightarrow +0$ limit.
By using the following two identities,
\begin{eqnarray}
\lim_{\epsilon \rightarrow +0} \log \epsilon 
(\exp(\frac{A}{\epsilon})+\exp(\frac{B}{\epsilon})) = \Max(A,B) ,\nonumber\\
\lim_{\epsilon \rightarrow +0} \log \epsilon 
(\exp(\frac{A}{\epsilon})\times\exp(\frac{B}{\epsilon})) = A+B ,
\end{eqnarray}
we can obtain the ultradiscrete KP equation from 
equation (\ref{discreteKP2}).
\begin{eqnarray}
Z_{q}+Z_{pr}+F_{q}+F_{pr} = 
\Max(Z_{p}+Z_{qr}+F_{p}+F_{qr}, Z_{r}+Z_{pq}+F_{r}+F_{pq})
\end{eqnarray}
As in the case of the discrete KP equation, this equation 
reduces to many soliton cellular automaton models.
See \cite{udkp}, for example.

The B\"acklund transformation equations for the ultradiscrete KP
equation are obtained by applying the similar procedure to 
the extended B\"acklund transformation equations.
Applying the transformation (\ref{ultradiscritization}) and
\begin{eqnarray}
g = \exp ( \frac{G}{\epsilon}),
\end{eqnarray}
and taking the $\epsilon \rightarrow +0$ limit,
we obtain the equations
\begin{eqnarray}
Z_{q}+F_{q}+G_{r}=\Max [Z_{r}+F_{r}+G_{q},Z_{qr}+F_{qr}+G] , \nonumber\\
Z_{p}+F_{p}+G_{r}=\Max [Z_{r}+F_{r}+G_{p},Z_{pr}+F_{pr}+G] , \nonumber\\
Z_{p}+F_{p}+G_{q}=\Max [Z_{q}+F_{q}+G_{p},Z_{pq}+F_{pq}+G] , \nonumber\\
Z_{pr}+F_{pr}+G_{q}=\Max [Z_{qr}+F_{qr}+G_{p},Z_{pq}+F_{pq}+G_{r}] .
\label{ultradiscrete_backlund}
\end{eqnarray}

The consistency condition for these equations is obtained by eliminating
the all variables from these equations.
Let us notice that we can eliminate $G_{r}$ and $G_{q}$, by substituting
second equation and third equation to first equation and fourth equation,
 respectively.
Then, we obtain the following two linear equations for two variables 
$G_{p}$ and $G$
\begin{eqnarray}
&&\Max(Z_{r}+Z_{q}+F_{r}+F_{q}+G_{p},Z_{q}+Z_{pr}+F_{q}+F_{pr}+G) \nonumber\\
&& \quad =\Max(Z_{r}+Z_{q}+F_{r}+F_{q}+G_{p}, 
\Max(Z_{r}+Z_{pq}+F_{r}+F_{pq},Z_{qr}+Z_{p}+F_{qr}+F_{p})+G)) , \nonumber\\
&& \Max(Z_{pr}+Z_{q}+F_{pr}+F_{q}+G_{p},
Z_{pr}+Z_{pq}+F_{pr}+F_{pq}+G)\nonumber\\
&& \quad =\Max(\Max(Z_{qr}+Z_{p}+F_{qr}+F_{p},Z_{pq}+Z_{r}+F_{pq}+F_{r})+G_{p},
Z_{pq}+Z_{pr}+F_{pq}+F_{pr}+G) . \nonumber\\
\end{eqnarray}
By applying the relation (\ref{consistency_two}), 
we can eliminate the rest of two variables.
After some calculation, we can obtain the equation for $F$
\begin{eqnarray}
&&\Max(Z_{q}+Z_{r}+Z_{pq}+Z_{pr}+F_{q}+F_{r}+F_{pq}+F_{pr},\nonumber\\
&& \qquad Z_{pr}+Z_{q}+F_{pr}+F_{q}+\Max(Z_{r}+Z_{pq}+F_{r}+F_{pq},
Z_{qr}+Z_{p}+F_{qr}+F_{p}), \nonumber\\
&& \qquad Z_{q}+Z_{pr}+F_{q}+F_{pr}+\Max(Z_{qr}+Z_{p}+F_{qr}+F_{p},
Z_{pq}+Z_{r}+F_{pq}+F_{r}))\nonumber\\
&&=\Max(Z_{q}+Z_{r}+Z_{pq}+Z_{pr}+F_{q}+F_{r}+F_{pq}+F_{pr}, \nonumber\\
&& \qquad 2(Z_{pr}+Z_{q}+F_{pr}+F_{q}),2\Max(Z_{r}+Z_{pq}+F_{r}+F_{pq},
Z_{qr}+Z_{p}+F_{qr}+F_{p})) .
\end{eqnarray}
Notice that we can remove the first terms of the both hand side,
since they are smaller than the other terms.
Moving the rest of the right hand side to the left hand side,
we obtain the equation
\begin{eqnarray}
\|
Z_{q}+Z_{pr}+F_{q}+F_{pr} 
-\Max(Z_{p}+Z_{qr}+F_{p}+F_{qr}, Z_{r}+Z_{pq}+F_{r}+F_{pq}
\| = 0 .
\end{eqnarray}
This is nothing but the absolute value of the ultradiscrete KP equation.

Since we can change the roles of $F$ and $G$, by shifting each of the equations  of (\ref{ultradiscrete_backlund}) appropriately,
$G$ should also satisfy the ultradiscrete KP equation.
Thus, we can show that 
equations (\ref{ultradiscrete_backlund}) really generate 
the B\"acklund transformation of the ultradiscrete KP equation.

In the rest of this section, we derive the two soliton solution
and conjecture the form of the N soliton solution.
Let us assume the existence of the trivial solution,
 $F=G=0$,
then the conditions for the coefficients of equations
(\ref{ultradiscrete_backlund}),
\begin{eqnarray}
Z_{q}=\Max [Z_{r},Z_{qr}] , \nonumber\\
Z_{p}=\Max [Z_{r},Z_{qr},Z_{pq}] , \nonumber\\
Z_{pr}=\Max [Z_{qr},Z_{pq}] ,
\end{eqnarray}
 should be satisfied.
Notice that $Z_{r},Z_{qr}$ and $Z_{pq}$ determine the rest of the
coefficients.
We rewrite these three coefficients as 
\begin{eqnarray}
Z_{pq}=Z_{1} \quad , \quad 
Z_{qr}=Z_{2} \quad , \quad
Z_{r}=Z_{3} \quad ,
\end{eqnarray}
for simplicity.
Furthermore we assume the relations $Z_{1} \le Z_{2} \le Z_{3}$,
then the rest of the coefficients becomes as 
\begin{eqnarray}
Z_{q}=Z_{2} \quad , \quad Z_{p}=Z_{1} \quad , \quad Z_{pr}=Z_{1} .
\end{eqnarray}
Of course we can consider the other relations for $Z_{1},Z_{2}$ and $Z_{3}$,
however, it does not change the following argument.

The one soliton solution is derived as follows.
First, we substitute the trivial solution $F=0$ into the
B\"acklund transformation equations (\ref{ultradiscrete_backlund}),
and assume the form of $G$ as
\begin{eqnarray}
G=\vec{\alpha} \cdot \vec{p} \label{pre_one_soliton} .
\end{eqnarray}
Here, $\vec{\alpha}$ is the three dimensional vector of constant coefficients,
and $\vec{p}$ is the vector of coordinates $p,q,r$
\begin{eqnarray}
\vec{\alpha}=(\alpha_{p},\alpha_{q},\alpha_{r}) \quad , \quad 
\vec{p}=(p,q,r) .
\end{eqnarray}
The dot between them represents the vector product.
By substituting the $G$ of equation (\ref{pre_one_soliton}) into
the B\"acklund transformation equations (\ref{ultradiscrete_backlund}),
we obtain the following dispersion relations
\begin{eqnarray}
Z_{2}+\alpha_{r}=\Max [Z_{3}+\alpha_{q},Z_{2}] , \nonumber\\
Z_{1}+\alpha_{r}=\Max [Z_{3}+\alpha_{p},Z_{1}] , \nonumber\\
Z_{1}+\alpha_{q}=\Max [Z_{2}+\alpha_{p},Z_{1}] , \nonumber\\
Z_{1}+\alpha_{q}=\Max [Z_{2}+\alpha_{p},Z_{1}+\alpha_{r}] .
\end{eqnarray}
We can solve these dispersion relations as
\begin{eqnarray}
\alpha_{p}=\alpha+Z_{1} \quad , \quad
\alpha_{q}=\Max(\alpha+Z_{2},0) \quad , \quad 
\alpha_{r}=\Max(\alpha+Z_{3},0) .
\end{eqnarray}
Here, $\alpha$ is an arbitrary real parameter.
The one soliton solution of the ultradiscrete KP equation becomes
the superposition of these representations of $G$ which is written in ``\Max'' 
operator,
\begin{eqnarray}
G=\Max(\vec{\alpha} \cdot \vec{p} \; ,\;
\vec{\beta} \cdot \vec{p}) .
\label{one_soliton}
\end{eqnarray}
where $\beta$ is a three dimensional vector $(\beta_{p},\beta_{q},\beta_{r})$
which satisfies the dispersion relations.

Figure 1 shows the time evolution of three soliton solution.
We transform the dependent variable $f$ and coordinates $p,q,r$ into $U$ and $l,m,n$ as
\begin{eqnarray}
U(p,q,r)=f(p,q,r+1)+f(p+1,q+1,r)-f(p+1,q,r)-f(p,q+1,r+1) , \nonumber\\
l=p-r \;,\; m=q \;,\; n=p+r .
\end{eqnarray}
The $l$ axis is parallel to the horizontal line and $m$ axis is parallel
to the vertical line.

We see the propagation of one soliton solution in the large $l$ and 
large $m$ region, in Fig. 1.

Similarly, we can obtain the two soliton solution 
by substituting the one soliton solution (\ref{one_soliton})
to the function $F$ in the B\"acklund transformation equations, and 
solve the equations for $G$.
Let us assume the form of $G$ as
\begin{eqnarray}
G=\Max(c_{\alpha}+\vec{\alpha}\cdot\vec{p}+\vec{\gamma}\cdot\vec{p} \; ,\;
        c_{\beta}+\vec{\beta}\cdot\vec{p}+\vec{\gamma}\cdot\vec{p}),
\label{pre_twosoliton}
\end{eqnarray}
where $c_{\alpha}$ and $c_{\beta}$ are the constants which should be
determined later, and $\vec{\gamma}$ is also a three dimensional vector
which satisfy the dispersion relations.
Substituting this expression into the B\"acklund transformation equations,
we obtain the following relations of $c_{\alpha}$ and $c_{\beta}$
\begin{eqnarray}
\left\{
\begin{array}{ccc}
c_{\alpha}=c_{\beta},\; & \mbox{if} \quad \gamma \le \alpha\; , \; \beta ,\\
c_{\alpha}-\alpha = c_{\beta}-\beta ,\;& \mbox{if} \quad \alpha\; ,\;\beta \le \gamma ,\\
\mbox{There is no solution of } c_{\alpha} \mbox{ and } c_{\beta} 
& \mbox{if} \quad \alpha \le \gamma \le \beta \quad \mbox{or} \quad 
\beta \le \gamma \le \alpha .
\end{array}\right .
\end{eqnarray}
These results can be summarized into the following forms, without the
loss of the generality
\begin{eqnarray}
&&c_{\alpha}=\Max(\alpha,\gamma) , \nonumber\\
&&c_{\beta}=\Max(\alpha,\gamma) \quad , \quad (\alpha \; ,\;\beta \le \gamma
\mbox{ or } \gamma \le \alpha \; ,\;\beta) .
\end{eqnarray}
The two soliton solution of the ultradiscrete KP equation
becomes the superposition of these solutions,
\begin{eqnarray}
G=\Max(\Max(\alpha,\gamma)+\vec{\alpha}\cdot\vec{p} +\vec{\gamma}\cdot\vec{p}
\;,\;
        \Max(\alpha,\delta)+\vec{\alpha}\cdot\vec{p}+\vec{\delta}\cdot\vec{p}
        \;,\nonumber\\
        \Max(\beta,\gamma)+\vec{\beta}\cdot\vec{p}+\vec{\gamma}\cdot\vec{p}
        \;,\;
        \Max(\beta,\gamma)+\vec{\beta}\cdot\vec{p}+\vec{\delta}\cdot\vec{p})
        \quad,
\end{eqnarray}
where $\gamma$ and $\delta$ should satisfy one of the following conditions
\begin{eqnarray}
\gamma \;,\;\delta \le \alpha\;,\;\beta \nonumber\\
\gamma \le \alpha\;,\;\beta \le \delta \nonumber\\
\alpha\;,\;\beta \le \gamma\;,\;\delta .
\end{eqnarray}

These arguments allow one to suppose the form of the N soliton solution.
The N soliton solution may have the following form
\begin{eqnarray}
f=\Max_{\{\sigma\}}(\sum_{i < j}\Max(\alpha_{i}^{\sigma_{i}},\alpha_{j}^{\sigma_{j}})
+\sum_{i}\vec{\alpha_{i}}^{\sigma_{i}}\cdot \vec{p}) .
\label{Nsoliton}
\end{eqnarray}
Here, $\{\sigma\}=(\sigma_{1},\sigma_{2},...,\sigma_{N})$ and
$\sigma_{i}$'s take value of 1 or 0.
$\Max_{\{\sigma\}}$ means to take the maximal value of all the combination
of the $\sigma$'s.
The vectors $\vec{\alpha_{i}}^{0}$ and $\vec{\alpha_{i}}^{1}$ are constant 
coefficients which satisfy the dispersion relations, 
and we also assume that one of the following conditions is
satisfied for all combinations of $i \le j$
\begin{eqnarray}
\alpha_{i}^{0} \;,\; \alpha_{i}^{1} \le 
\alpha_{j}^{0} \;,\; \alpha_{j}^{1} , \nonumber\\
\alpha_{i}^{0} \le \alpha_{j}^{0} \;,\; \alpha_{j}^{1} \le \alpha_{j}^{1} ,\nonumber\\
\alpha_{j}^{0} \;,\; \alpha_{j}^{1} \le \alpha_{i}^{0} \;,\; \alpha_{i}^{1} .
\end{eqnarray}
We do not have the proof of the expression (\ref{Nsoliton}), 
but we confirm that it really satisfies 
the B\"acklund transformation equations
till $N$ = 4.
The complete proof will be given in the other paper.

\begin{figure}
\includegraphics[scale=0.9]{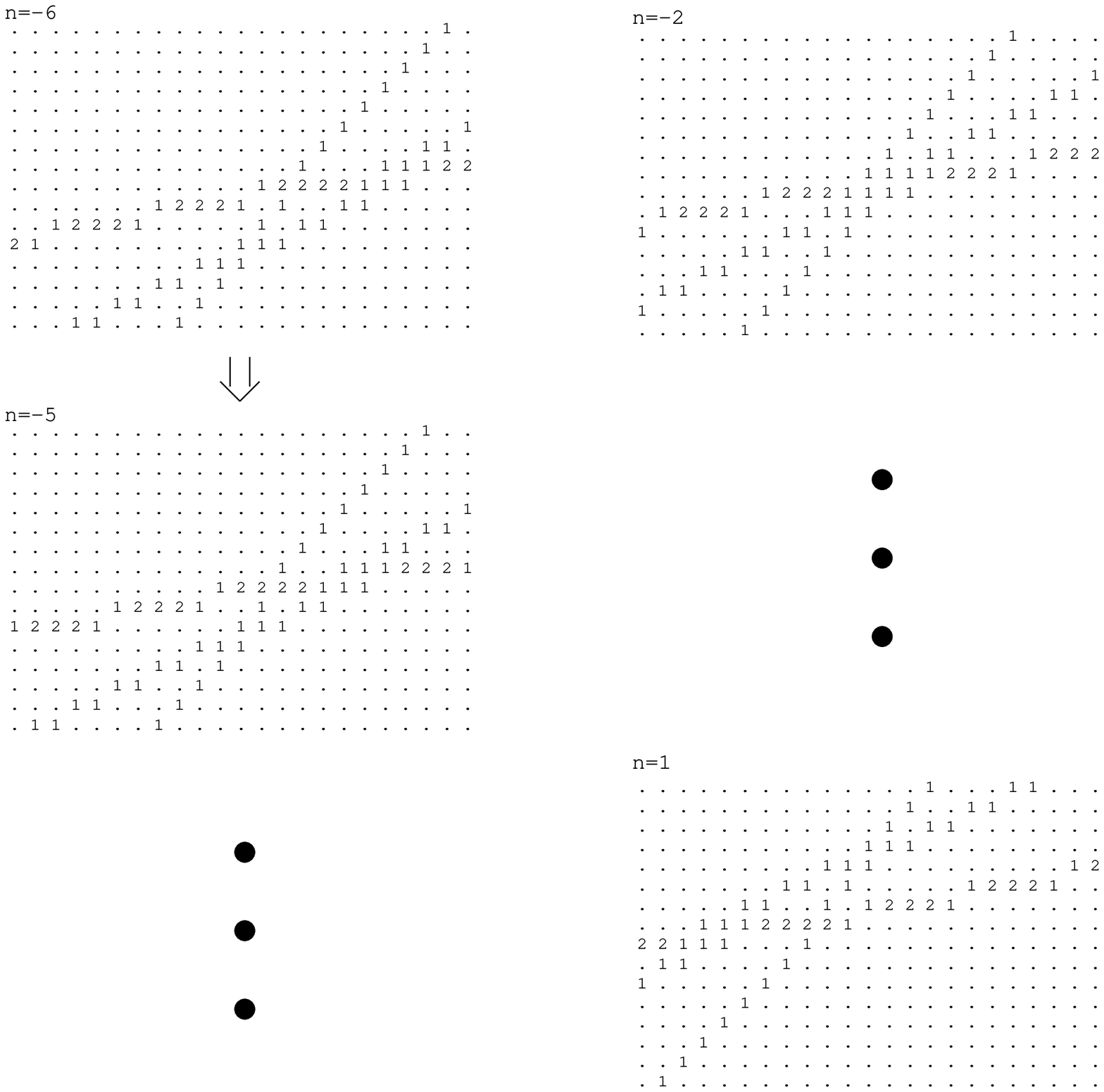}
\begin{eqnarray*}
Z_{1}=0 \;,\; Z_{2}=1 \;,\; Z_{3}=0 ,\nonumber\\
\alpha_{1}^{0}=0 \;,\; \alpha_{1}^{1}=-1 ,\nonumber \\
\alpha_{2}^{0}=1 \;,\; \alpha_{2}^{1}=-1 ,\nonumber \\
\alpha_{3}^{0}=1 \;,\; \alpha_{3}^{1}=-4 .
\end{eqnarray*}
\caption{Time evolution of 3 soliton solution}
\end{figure}

\section{Conclusion}
We consider the consistency condition for the B\"acklund transformation 
equations of the ultradiscrete KP equation.
We extend the B\"acklund transformation equations for the discrete KP equation
into the form of the product of a matrix and a vector.
The consistency condition becomes the vanishing of the determinant 
of the matrix.
In this case, the matrix becomes an antisymmetric matrix, and we can 
obtain the discrete KP equation by using the relation
between the determinant of an anti symmetric matrix and a pfaffian. 
We also show the algorithm to eliminate the variables
from the ultradiscrete linear equations.
By using this algorithm, we obtain the consistency condition
of the B\"acklund transformation equations of the ultradiscrete KP equation.
The consistency condition becomes the
absolute value of the ultradiscrete KP equation, instead of 
the square of the pfaffian in the discrete KP equation case.

\end{document}